\documentclass[aps,prc,nofootinbib,preprint,groupedaddress]{revtex4-1}

\usepackage[normalem]{ulem}  
\usepackage[dvipdfmx]{color} 
\usepackage{amsmath,amssymb}
\usepackage[dvipdfmx]{graphicx}
\usepackage{float}
\usepackage{wrapfig}
\newcommand{\Slash}[1]{{\ooalign{\hfil/\hfil\crcr$#1$}}}
\newcommand{\tr}{{\rm tr}}

\newcommand{\etaN}{{\eta N}}

\newcommand{\diag}{{\rm diag}}
\newcommand{\tepep}{T_{\eta'N\rightarrow\eta'N}}
\newcommand{\tetacm}{\theta_{\eta'}^{\rm c.m.}}

\renewcommand\sout{\bgroup \color[rgb]{1,0,0} \ULdepth=-.5ex \ULset}
\allowdisplaybreaks[3]

\begin{document}


\title{Effect of the final state interaction of $\eta' N$ on the $\eta'$
photoproduction off the nucleon}


\author{Shuntaro
Sakai$^1$\footnote{shsakai@rcnp.osaka-u.ac.jp\\Present address:
Departamento de F\'isica Te\'orica and IFIC, Centro Mixto Universidad de
Valencia-CSIC, Institutos de Investigaci\'on de Paterna, Aptdo. 22085,
46071 Valencia, Spain}}
\author{Atsushi Hosaka$^{1,2}$}

\author{Hideko Nagahiro$^{1,3}$}
\affiliation{$^1$Research Center for Nuclear Physics (RCNP), Osaka
University, Ibaraki, Osaka, 567-0047, Japan.\\
$^2$Advanced Science Research Center, Japan Atomic Energy Agency, Tokai,
Ibaraki, 319-1195, Japan.\\
$^3$Department of Physics, Nara Womens's University, Nara
630-8506, Japan.}


\date{\today}

\begin{abstract}
 We investigate the $\eta'$ photoproduction off the nucleon with a
 particular interest in the effect of the final-state interaction (FSI)
 of the $\eta'$ meson and nucleon $(\eta'N)$ based on the three-flavor
 linear $\sigma$ model.
 We find an enhancement in the cross section of the $\eta'$
 photoproduction near the $\eta'N$-threshold energy owing to
 the $\eta'N$ FSI.
 With the $\eta'$ meson at forward angles,
 the energy dependence near the $\eta'N$ threshold is well
 reproduced with the $\eta'N$ FSI.
 The cross section at backward angles can also be a good probe to
 investigate the strength of the $\eta'N$ interaction.
\end{abstract}


\maketitle


\section{Introduction\label{sec_intro}}
Hadrons are elementary excitations of the vacuum of quantum
chromodynamics (QCD),
and their properties reflect the vacuum structure of the low-energy QCD.
The chiral symmetry is a basic feature of QCD, and it is broken
spontaneously at low energies.
In the nuclear medium,
the spontaneously broken symmetry is expected to be restored,
which we call chiral restoration, and
the possible change of hadron properties
associated with the chiral restoration
at finite baryon densities has been an important subject of hadron
physics (see, for example, Ref.~\cite{Hayano:2008vn} for a recent review
article).
For example, some theoretical analyses suggest the mass reduction of
vector mesons as an evidence of the restoration of the chiral symmetry
\cite{Brown:1991kk,*Hatsuda:1991ez}.
There exist some attempts and discussions for the study of the in-medium
properties of the vector meson from the theoretical and experimental
sides~\cite{Naruki:2005kd,*Friedrich:2014lba,*Gubler:2016itj}.
In the case of the $\omega$ meson, the experimental data are consistent
with the weakly attractive optical potential.
The analyses of the pion-nucleus system suggest
partial restoration of chiral symmetry in the nuclear medium;
the pion decay constant, the order parameter of the spontaneous breaking
of chiral symmetry, is expected to be reduced about 35\% at the normal
nuclear density
\cite{Suzuki:2002ae,Friedman:2004jh,Kolomeitsev:2002gc,Jido:2008bk}.

The pseudoscalar meson $\eta'$ is another candidate to probe such a
change of the vacuum property.
Its mass is larger than other low-lying pseudoscalar mesons, such as $\pi$,
$K$, or $\eta$, due to the chiral symmetry breaking in the three-flavor
system
\cite{Pisarski:1983ms,Kikuchi:1987jr,Kunihiro:1987bb,*Kunihiro:1989my,Cohen:1996ng,Lee:1996zy,Evans:1996wf,Birse:1996dx}
together with the U$_A$(1) anomaly in QCD
\cite{Bardeen:1969md,*Kobayashi:1970ji,*Schechter:1971qa,*'tHooft:1976fv,*Witten:1979vv,*Veneziano:1979ec,*Kawarabayashi:1980dp,*Rosenzweig:1979ay}.
According to the argument of the restoration of the chiral symmetry in
nuclear medium, the $\eta$-$\eta'$ mass difference can be as large as
150 MeV at the normal nuclear density,
even if the property of the U$_A$(1) anomaly is unchanged in the
medium~\cite{Jido:2011pq}.
So far there are many theoretical and experimental studies
both for the $\eta$ and $\eta'$ to investigate property changes of these mesons
\cite{Pisarski:1983ms,Kikuchi:1987jr,Kunihiro:1989my,Cohen:1996ng,Lee:1996zy,Evans:1996wf,Birse:1996dx,Jido:2011pq,Kapusta:1995ww,Costa:2002gk,Bardeen:1969md,*Kobayashi:1970ji,*Schechter:1971qa,*'tHooft:1976fv,*Witten:1979vv,*Veneziano:1979ec,*Kawarabayashi:1980dp,*Rosenzweig:1979ay,Nagahiro:2006dr,Nagahiro:2004qz,Bass:2005hn,Saito:2005rv,Kwon:2012vb,Sakai:2013nba,Nagahiro:2011fi,Miyatani:2016pxz,*Csorgo:2009pa,*Benic:2011fv,*Fejos:2016hbp,Pfeiffer:2003zd,*Mersmann:2007gw,*Smyrski:2007nu,*Pheron:2012aj,Waas:1997pe,*Jido:2008ng,*Nagahiro:2008rj,Nanova:2012vw,Itahashi:2012ut,Nanova:2013fxl,Nanova:2016cyn,Tanaka:2016bcp}.

For the $\eta'$ meson, several interesting experimental results have
been recently reported and/or planned
\cite{Nanova:2012vw,Itahashi:2012ut,Nanova:2013fxl,Nanova:2016cyn,Tanaka:2016bcp,Moyssides:1983,Moskal:2000gj,*Moskal:2000pu,*Czerwinski:2014yot,Williams:2009yj,Sumihama:2009gf,Crede:2009zzb,Morino:2013raa,Sandri:2014nqz,Kashevarov:2017kqb}
which give the information on the $\eta'N$ interaction and
$\eta'$-nucleus interaction.
These possible property changes of the $\eta'$ in nuclear medium have
been also discussed at several kinematical situations.
Unfortunately, so far there is no theoretical framework to explain all
the available data consistently.
One of the reasons is in the complexity coming from the nuclear many
body effects for mesons in nuclear medium, and hence in the extraction
of the basic hadron interactions.
Therefore, comparisons between theoretical predictions and experimental
observables are not so simple.
An example is the $\eta$-nucleus system
\cite{Waas:1997pe,*Jido:2008ng,*Nagahiro:2008rj,Pfeiffer:2003zd,*Mersmann:2007gw,*Smyrski:2007nu,*Pheron:2012aj}.
From a naive chiral symmetry argument, the mass of $\eta$ meson does not
change much in nuclear medium because of the Nambu-Goldstone nature.
However, it is also known that the $\eta$ meson couples strongly with
the $N^\ast(1535)$, which in nuclear medium provides strong attraction.
This attraction for $\eta$ is also referred to be as effective mass
reduction of $\eta$, which should be different from that due to partial
restoration of chiral symmetry.

In such situations, we consider it very important to know the basic
interactions of relevant mesons and nucleons{, which is investigated theoretically in
Refs.~\cite{Kawarabayashi:1980uh,Bass:1999is,Borasoy:1999nd,Oset:2010ub,Sakai:2014zoa,Sekihara:2016jgx,Kashevarov:2016owq,Zhang:1995uha,Li:1996wj,Borasoy:2001pj,Bass:2000np,Borasoy:2002mt,Chiang:2002vq,Sibirtsev:2003ng,Nakayama:2004ek,Nakayama:2005ts,Tryasuchev:2008zz,Cao:2008st,Zhong:2011ht,Huang:2012xj}.}
To this end, in the present paper we investigate the $\eta'$
photoproduction off a free nucleon
with the final-state interaction (FSI) between the $\eta'$ meson
and nucleon,
which is the simplest process for the $\eta'N$ interaction.
{For this purpose}, we employ a three-flavor linear $\sigma$ model.
{In this model, a} strongly attractive $\eta'N$ interaction is
allowed due to the U$_A$(1) anomaly and the scalar-meson exchange,
such that an $\eta'N$ bound state is generated with a
binding energy of typically about a few tens of MeV~\cite{Sakai:2013nba,Sakai:2014zoa}.
In the present study, we supplement the $\rho$ meson for the $\eta'$
photoproduction, which is empirically known to be important,
in the linear $\sigma$ model with relevant couplings fixed by existing data.
We then focus on the final-state interaction of the $\eta'$ meson
with the nucleon which can affect the energy dependence of the production
cross sections near and above the threshold.
For this purpose, we perform our analysis by changing the strength
of the $\eta'$-nucleon coupling from the original one of the linear
$\sigma$ model.
By doing this, we discuss how the effect of the $\eta'$-$N$ interaction
shows up in the observed experiment.

This paper is organized as follows.
In Sec.~\ref{sec_mpdel_setup}, we explain the model setup used in this
analysis of the $\eta'$ photoproduction.
The $\eta'N$ interaction and the photoproduction amplitude
used in the present study are also explained in this section.
Section~\ref{sec_result} is devoted to the discussion of the
cross section and the beam asymmetry of the $\eta'$ photoproduction off
the nucleon with the inclusion of the $\eta'N$ FSI.
The summary and outlook of this study is given in Sec.~\ref{sec_summary}.

\section{Formulation 
\label{sec_mpdel_setup}}

\subsection{Model Lagrangian
\label{subsec_lagrangian}}
In this section, we explain the model setup for the $\eta'$
photoproduction in the three-flavor linear $\sigma$ model.
In the linear model hadrons including pseudoscalar mesons, scalar
mesons, and baryons are introduced as linear representations of chiral
symmetry and their interactions are determined.
This is done first by constructing a chiral invariant Lagrangian and then
the vacuum is determined to minimize the effective potential;
the neutral scalar fields have non-zero expectation values in
association with the chiral symmetry breaking.
Hadron properties in such a framework can naturally be related to the
vacuum structure.

The Lagrangian used in this calculation is given by
\begin{align}
 \mathcal{L}=&\mathcal{L}_{M}+\mathcal{L}_{N}+\mathcal{L}_{\gamma VP},\label{eq_nc_vct_cplng}\\
 \mathcal{L}_{M}=&\frac{1}{2}\tr\left[D_\mu M(D^\mu
 M)^\dagger\right]-\frac{\mu^2}{2}\tr\left[MM^\dagger\right]-\frac{\lambda}{4}\tr\left[(MM^\dagger)^2\right]-\frac{\lambda'}{4}\left[\tr\left(MM^\dagger\right)\right]^2\notag\notag\\
 &+\sqrt{3}B(\det
 M+\det M^\dagger)+A\tr\left(\chi
 M^\dagger+M\chi^\dagger\right)\notag\\
 &-\frac{1}{4}\tr\left[(L^{\mu\nu})^2+(R^{\mu\nu})^2\right]+\frac{m_0^2}{2}\tr\left[(L^\mu)^2+(R^\mu)^2\right],\notag\\
 \mathcal{L}_{N}=&\bar{N}\left[
 i\left\{\Slash{\partial}+ig_V\left(\Slash{V}+\frac{\kappa_V}{2m_N}\sigma_{\mu\nu}\partial^\mu
 V^\nu\right)\right\}\right.\notag\\
 &\left.\hspace{5mm}-m_N-g\left\{\left(\frac{\tilde{\sigma}_0}{\sqrt{3}}+\frac{\tilde{\sigma}_8}{\sqrt{6}}\right)+i\gamma_5\left(\frac{\eta_0}{\sqrt{3}}+\frac{\vec{\pi}\cdot\vec{\tau}}{\sqrt{2}}+\frac{\eta_8}{\sqrt{6}}\right)\right\}\right]N\notag\\
 \mathcal{L}_{\gamma VP}=&eg_{\gamma
 V^aP}\epsilon^{\mu\nu\alpha\beta}(\partial_\mu V^a_\nu)
 (\partial_\alpha A_{{\rm em}\beta})\eta',\notag\\
 &D_\mu M=\partial_\mu M+ig_V(L_\mu M-MR^\dagger_\mu),\notag\\ 
 &M=M_s+iM_{ps}=\sum_{a=0}^8\frac{\sigma^a\lambda^a}{\sqrt{2}}+i\sum_{a=0}^8\frac{\pi^a\lambda^a}{\sqrt{2}},\notag\\
 &N=^t(p,n),V^\mu=\frac{1}{\sqrt{2}}\sum_{a=0}^3\frac{V^{a\mu}\tau^a}{\sqrt{2}},\notag\\
 &\chi=\sqrt{3}{\diag}(m_u,m_d,m_s)= \sqrt{3}{\diag}(m_q,m_q,m_s),\notag
\end{align}
where we write $L^\mu$ and $R^\mu$ as $L^\mu=V^\mu+A^\mu$ and
$R^\mu=V^\mu-A^\mu$ using the vector and the axial-vector fields $V^\mu$
and $A^\mu$, and $e>0$ is the elementary charge unit.
$A_{\rm em}^\mu$ denotes the electromagnetic field.
$\tilde{\sigma}_i$ $(i=0,8)$ appearing in the nucleon part is the
fluctuation of the neutral scalar field from its mean field.
The mean field is determined so as to minimize the effective potential,
which is obtained in the tree-level approximation in this study.
The isospin symmetry is implemented with the degenerate $u$ and $d$
quark masses.
The Lagrangian except for the vector field is the same as that used in
Refs.~\cite{Sakai:2013nba,Sakai:2014zoa}.

The Lagrangian is constructed to be invariant under the chiral
transformation for the hadron field.
The meson field $M$ is transformed as $U_LMU_R^\dagger$ with $U_{L/R}$
the element of SU(3)$_{L/R}$.
Here, we note that the term proportional to $B$, which is not invariant
under the U$_A$(1) transformation, reflects the effect of the U$_A$(1)
anomaly.
For the fermion part, the irrelevant hyperons in this study are omitted.
The values of various coupling constants in the Lagrangian are
summarized in Table~\ref{tab_coup_val}.
The coupling of the vector meson and nucleon $g_V$ is fixed with the
Kawarabayashi-Suzuki-Fayyazuddin-Riazuddin relation
$g_V=\frac{m_V}{\sqrt{2}f}$
\cite{Kawarabayashi:1966kd,*Fayyazuddin:1962gnz},
where $m_V$ and $f$ are
$m_V=(m_\rho+m_\omega)/2$ and $f=92.2$ MeV.
The masses of the $\rho$ and $\omega$ mesons are
taken from Ref.~\cite{Agashe:2014kda}.
The coefficient of the Pauli coupling between the nucleon and vector
meson $\kappa_V$ is determined to reproduce the anomalous magnetic
moment of proton, $\kappa_p=1.793$.
Following Refs.~\cite{Sakai:2013nba,Sakai:2014zoa}, the parameter $g$
in the nucleon part is determined for $\left<\sigma\right>$,
the chiral order parameter, to reduce 35\% at the normal nuclear density
which is suggested by
the analysis of the pion-nucleus system \cite{Suzuki:2002ae}.
For the masses of the $\eta$, $\eta'$ mesons and the nucleon,
there are constraints in our model.
However, in the present study of the $\eta'$ photoproduction, we employ
the experimental values for these masses.
The coupling of the photon $\gamma$, vector meson $V^a$ ($V^0=\omega$
and $V^3=\rho^0$), and the $\eta'$ meson is called the anomalous
coupling which is induced by the chiral anomaly in QED
\cite{Adler:1969gk,*Bell:1969ts} with the vector meson dominance.
Here, we use $g_{\gamma V^a\eta'}$
determined from the observed partial width of the $\eta'$ radiative
decay \cite{Agashe:2014kda}.
\begin{table}[t]
 \centering
 \caption{Values of the parameters in the Lagrangian.
 \label{tab_coup_val}}
 \begin{ruledtabular}    
 \begin{tabular}[t]{cccccc}
  $g_V[-]$&$\kappa_\rho[-]$ &$\kappa_\omega[-]$&$g_{\gamma\eta'\rho}[{\rm
	      MeV}^{-1}]$&$g_{\gamma\eta'\omega}[{\rm MeV}^{-1}]$&$g[-]$ \\\hline
  $5.95$&$3.586$ &$0$ &$1.625\times 10^{-3}$&$5.622\times10^{-4}$&7.698
 \end{tabular}
 \end{ruledtabular}
\end{table}

\subsection{$\eta'N$ amplitude for FSI
\label{subsec_epN_int}}
\begin{figure}[t]
 \centering
 \includegraphics[width=15cm]{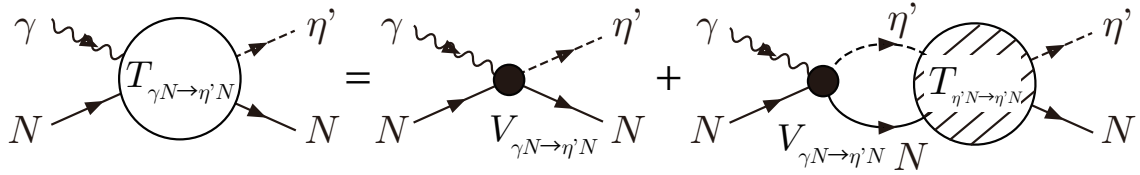}
 \caption{Diagrams for the scattering equation Eq.~(\ref{eq_t_prod}) for the
 $\eta'$-photoproduction $T$ matrix.
 The white, small black, and shaded blobs are the
 $T$ matrix of the $\eta'$ photoproduction, $T_{\gamma
 N\rightarrow\eta'N}$, the $\eta'$-photoproduction
 vertex, $V_{\gamma N\rightarrow\eta'N}$, and the $\eta'N$ $T$ matrix,
 $\tepep$.
 The solid, dashed, and wavy lines stand for the propagations of the
 nucleon, $\eta'$ meson, and photon, respectively.}
 \label{fig_t_prod}
\end{figure}

\begin{figure}[t]
 \centering 
 \includegraphics[width=15cm]{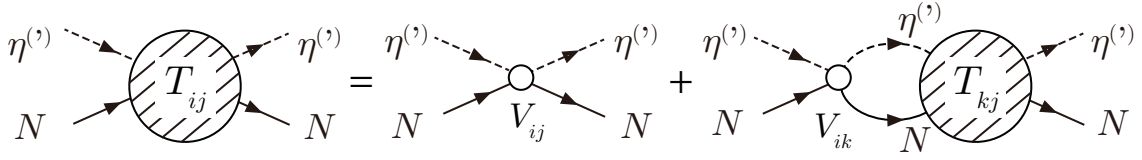}
 \caption{A diagrammatic representation of the scattering matrix for the
 $\eta'N$ scattering including the $\eta'N$ and $\eta N$ channels,
 $T_{ij}$ $(i,j=\eta'N,\etaN)$.
 The shaded and small white blobs represent the $T$ matrices, $T_{ij}$,
 and interaction kernels, $V_{ij}$.
 The solid line denotes the propagation of the nucleon,
 and the dashed one represents that of the $\eta$ or $\eta'$ meson.}
 \label{fig_epn_scat}
\end{figure}

\begin{figure}[t]
 \includegraphics[width=15cm]{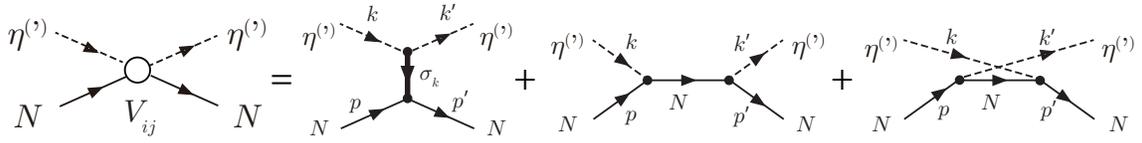}
 \caption{Diagrams for the interaction kernels $V_{ij}$ in
 Eq.~(\ref{eq_epn_scat}) at the tree level.
 The thick solid line stands for the propagation of the scalar meson, and
 the other lines have the same notations as those in
 Fig.~\ref{fig_epn_scat}.
 The first term is the contribution from the scalar-meson exchange in
 the $t$ channel, and
 the second and third ones represent the Born terms in the $s$ and
 $u$ channels, respectively.}
 \label{fig_tree_diag_etaprimeN}
\end{figure}

In this section, we briefly revisit the $\eta'N$ amplitude in the
framework of the linear $\sigma$ model \cite{Sakai:2013nba,Sakai:2014zoa},
which is
relevant
to the purpose of this study.
The $\eta'$ photoproduction amplitude is given by the
$T$ matrix $T_{\gamma N\rightarrow\eta'N}$ as
\begin{align}
 T_{\gamma N\rightarrow\eta'N}=V_{\gamma N\rightarrow\eta'N}(1+G_{\eta'N}\tepep),\label{eq_t_prod}
\end{align}
whose diagrammatic expression is shown in Fig.~\ref{fig_t_prod}.
In Eq.~(\ref{eq_t_prod}), $V_{\gamma N\rightarrow\eta'N}$, $G_{\eta'N}$, and
$\tepep$ are the $\eta'$-photoproduction kernel, the $\eta'N$
two-body Green's function, and the $\eta'N$ $T$ matrix,
respectively.
The amplitude $\tepep$ is responsible for the rescattering of the $\eta'$ meson and nucleon in the final state.
The $T$ matrix
is obtained from a two-channel coupled
equation of $\eta'N$ $(i=1)$ and $\eta N$ $(i=2)$.
With the interaction kernels of the
$\eta'N$ and $\etaN$ channels $V_{ij}$ $(i,j=
1,2)$, the $T$ matrices $T_{ij}$ satisfy the scattering equation given
by,
\begin{align}
 T_{ij}=V_{ij}+V_{ik}G_kT_{kj},\label{eq_epn_scat}
\end{align}
where
\begin{align}
 V_{11}=-\frac{6gB}{\sqrt{3}m_{\sigma_0}^2},V_{12}=V_{21}=+\frac{6gB}{\sqrt{6}m_{\sigma_8}^2},V_{22}=0.\label{eq_eprimeN_kernel}
\end{align}
The diagrammatic expression of Eq.~(\ref{eq_epn_scat}) is given in
Fig.~\ref{fig_epn_scat}.
The interaction kernels $V_{ij}$ given in
Eq.~(\ref{eq_eprimeN_kernel}) are obtained from
the scattering amplitude within the
tree-level approximation and the leading order of the momentum expansion
in the flavor SU(3) symmetric limit.
The diagrams taken into account in this calculation are shown in
Fig.~\ref{fig_tree_diag_etaprimeN},
where
the scalar-meson exchange in the $t$ channel and the
Born diagrams in the $s$ and $u$ channels are considered.
One can see in Eq.~(\ref{eq_eprimeN_kernel}) that
an attractive interaction between the $\eta'$ meson and nucleon
is induced by the scalar-meson exchange in this approximation.
It is noteworthy that this interaction kernel is proportional to $B$,
which reflects the effect of the U$_A$(1) anomaly
as we mentioned in Sec.~\ref{subsec_lagrangian}.
Owing to this attraction, the bound state of the $\eta'$ meson and
nucleon can be generated.
In this study, the vector-meson contribution for the $\eta'N$
interaction is not taken into account, because
they do not give the leading contribution in the momentum expansion.
Here, we take account of the $\eta'N$ and $\etaN$ channels, and omit
the $\pi N$ one,
because we expect that the contribution from
the $\pi N$ channel would be small owing to the smallness of the $\pi
N\rightarrow\eta'N$ cross section \cite{Rader:1973mx}.
The divergence contained in the two-body Green's function $G_i$ is
removed by the dimensional regularization and the subtraction
constant is fixed with the natural renormalization scheme
\cite{Hyodo:2008xr}.
The relevant parameters are given as
$B=997.95$ MeV, $g=7.698$, $m_{\sigma_0}=700$ MeV, $m_{\sigma_8}=1225$
MeV, and the subtraction constants are $a_{\eta'N}(\mu=m_N)=-1.838$ and
$a_{\eta N}(\mu=m_N)=-1.239$, where $\mu$ is the renormalization point.
We use the same subtraction constant appearing in the Green's
function in Eq.~(\ref{eq_t_prod}) as that in the $\eta'N$ $T$ matrix in
Eq.~(\ref{eq_epn_scat}).

For the purpose of seeing the effect of the $\eta'N$ FSI,
we show the result with varying the parameter $g$
as (a) $g\times 0.0$, (b) $1.0$, (c) $0.5$, (d) $1.5$, and (e)
$-0.5$.
The coupling strength, $\eta'N$ scattering length, and binding
energy of the $\eta'N$ bound state in these cases are summarized in
Table~\ref{tab_epn_scat}.
\begin{table}[t]
 \centering
 \caption{Table for the coupling strength, scattering length, and
 binding energy for the cases (a) to (e).
 The cases (a) to (e) are characterized by the coupling parameter $g$.
 \label{tab_epn_scat}}
 \label{tab_strength}
 \begin{ruledtabular}
  \begin{tabular}[t]{c|ccccc}
   &(a)&(b)&(c)&(d)&(e)\\\hline
   coupling strength &$g\times 0.0$ &$g\times 1.0$ &$g\times 0.5$
	       &$g\times 1.5$ &$g\times -0.5$ \\
   scattering length [fm]  &$0.0$ &$-1.9+0.58i$ &$+0.53+0.042i$
	       &$-0.77+0.086i$ &$-0.13+2.8\times 10^{-3}i$ \\
   binding energy [MeV]  &$-$ &$9.79-7.10i$ &$-$ &$98.6-24.6i$ &$-$ \\
  \end{tabular}
 \end{ruledtabular}
\end{table}
In this model, there is no parameter set reproducing the scattering
length
suggested by the COSY-11 experiment
\cite{Czerwinski:2014yot}, which has the larger imaginary part than
the real part.
On the other hand, the $\eta'$ optical potential by CBELSA/TAPS seems to
be consistent with {scattering length} of case (c) in
Table~\ref{tab_strength} within the errors by the use of the
linear density approximation though it might be a
crude one.
Here, we do not restrict our analysis to the scattering length
suggested by the analysis of COSY-11~\cite{Moskal:2000gj} for an
independent and complementary analysis of the $\eta'N$ interaction.

\subsection{Photoproduction amplitude
\label{subsec_photopro_tree}}
\begin{figure}[t]
 \centering
 \includegraphics[width=15cm]{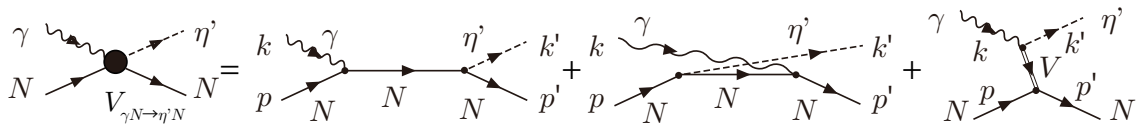}
 \caption{Diagrams for the $\eta'$ photoproduction amplitude
 with the tree-level approximation, with the same notation as in
 Fig.~\ref{fig_t_prod}.
 The first, second, and third terms are the contributions
 from the $s$, $u$, and $t$ channels, respectively.}
 \label{fig_diag_tree_photopro}
\end{figure}

In this section, we explain the $\eta'$-photoproduction
kernel $V_{\gamma
N\rightarrow\eta'N}$ in Eq.~(\ref{eq_t_prod}).
It is evaluated within the tree-level approximation shown in
Fig.~\ref{fig_diag_tree_photopro}, which contains the Born diagrams in
the $s$ and $u$ channels, and the vector-meson exchange one in the $t$
channel.
We can write down the amplitude from these diagrams as follows:
\begin{align}
 -i\mathcal{M}_{\rm
 tree}=&e\bar{u}(p',s')\left[g_{PN}\left\{\gamma_5\frac{F_s\Slash{k}+F_c(\Slash{p}+m_N)}{(p+k)^2-m_N^2}\Slash{\epsilon}+\Slash{\epsilon}\frac{-F_u\Slash{k}'+F_c(\Slash{p}+m_N)}{(p-k')^2-m_N^2}\gamma_5\right.\right.\notag\\
 &\left.+\frac{\kappa_p}{4m_N}\left(F_s\gamma_5\frac{\Slash{p}+\Slash{k}+m_N}{(p+k)^2-m_N^2}\left[\Slash{k},\Slash{\epsilon}\right]+F_u\left[\Slash{k},\Slash{\epsilon}\right]\frac{\Slash{p}-\Slash{k}'+m_N}{(p-k')^2-m_N^2}\gamma_5\right)\right\}\notag\\ 
 &\left.+iF_t\frac{
 g_Vg_{\gamma V^aP}/2}{t-m_V^2+i\epsilon}g_{\mu\sigma}\epsilon^{\rho\sigma\alpha\beta}k'_\rho k_\alpha\epsilon_\beta\left\{\gamma^\mu+\frac{\kappa_{V^a}}{4m_N}[\Slash{q},\gamma^\mu]\right\}\right]u(p,s),\label{eq_mat_elem_prod}
\end{align}
where $\epsilon^\mu$ is the polarization vector of the photon, and the
momentum transfer $q^\mu$ is written as $q^\mu=p'^\mu-p^\mu$.
Here, the form factors $F_x$ $(x=s,t,u)$ are introduced
in a gauge invariant manner
following Ref.~\cite{Choi:2005ki,*Choi:2007gy} and references therein.
The form factors $F_x$ appearing in Eq.~(\ref{eq_mat_elem_prod})
are written as
$F_x=\Lambda_x^4/((x-m_x^2)^2+\Lambda_x^4)$,
and $F_c$ is given as $F_c=F_s+F_u-F_sF_u$,
where $m_x$ denotes the exchanged hadron mass in the channel $x$.
The form factor reflects the size of hadron, and the typical value of
the cutoff parameter $\Lambda_x$ is
about $1$ GeV.
We will discuss the actual values in the next section.
For the kernel $V_{\gamma N\rightarrow\eta'N}$ in Eq.~(\ref{eq_t_prod}),
we use the production amplitude $\mathcal{M}_{\rm tree}$ in
Eq.~(\ref{eq_mat_elem_prod}) by factorizing the amplitude with its
on-shell value.

In the present calculation, we have omitted
the direct production of the $\eta$ meson
from the photon $V_{\gamma N\rightarrow\etaN}$,
expecting that the energy dependence of that
channel is not very large in the region of the $\eta'N$ threshold
because the pole position of $N^\ast(1535)$, which has the dominant
contribution for the $\eta$-meson photoproduction, is
far from there.

\section{Result
\label{sec_result}}
\begin{figure}[t]
 \begin{minipage}[t]{0.49\hsize}
  \centering
  \includegraphics[width=8cm]{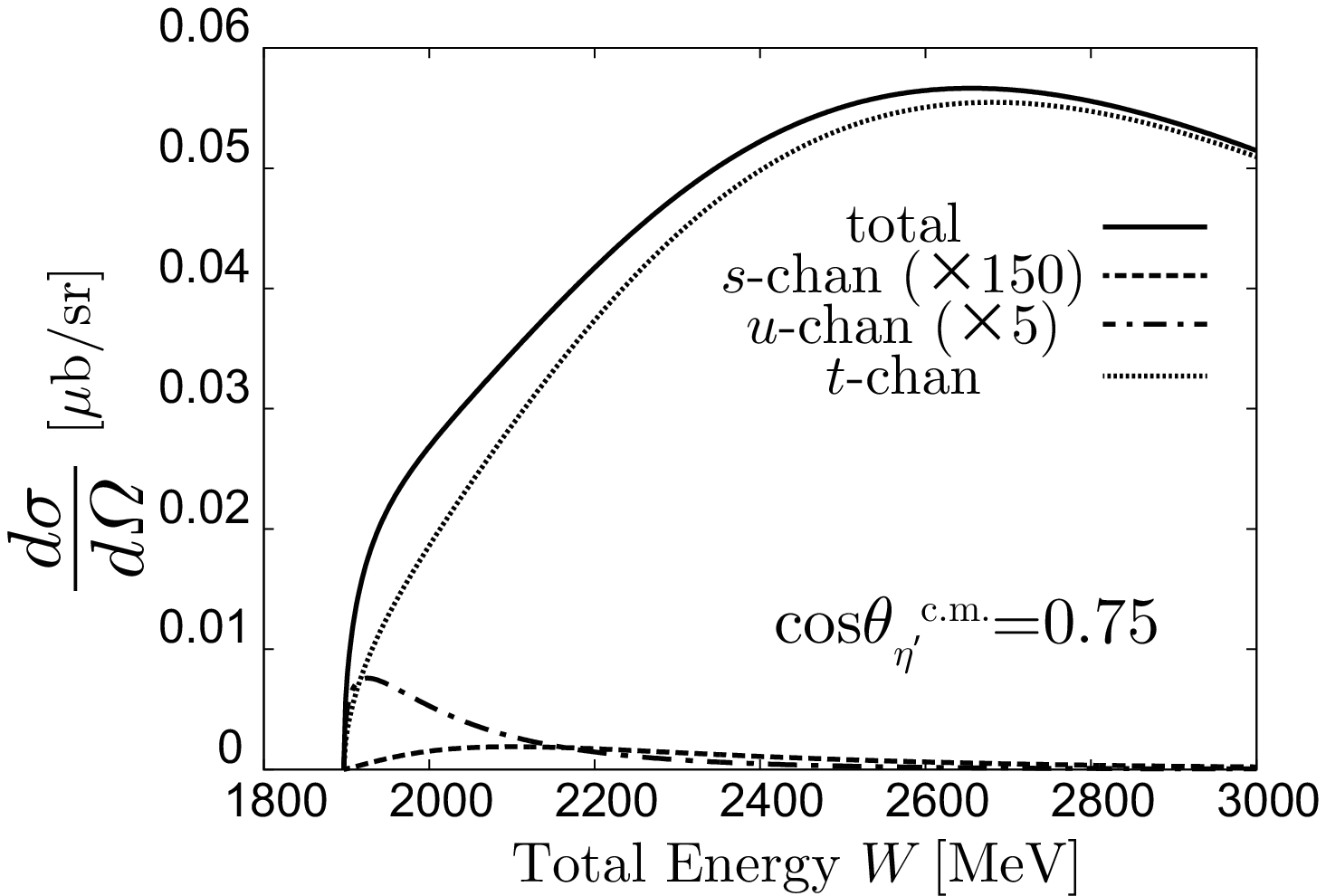}
 \end{minipage}
 \begin{minipage}[t]{0.49\hsize}
  \centering 
  \includegraphics[width=8cm]{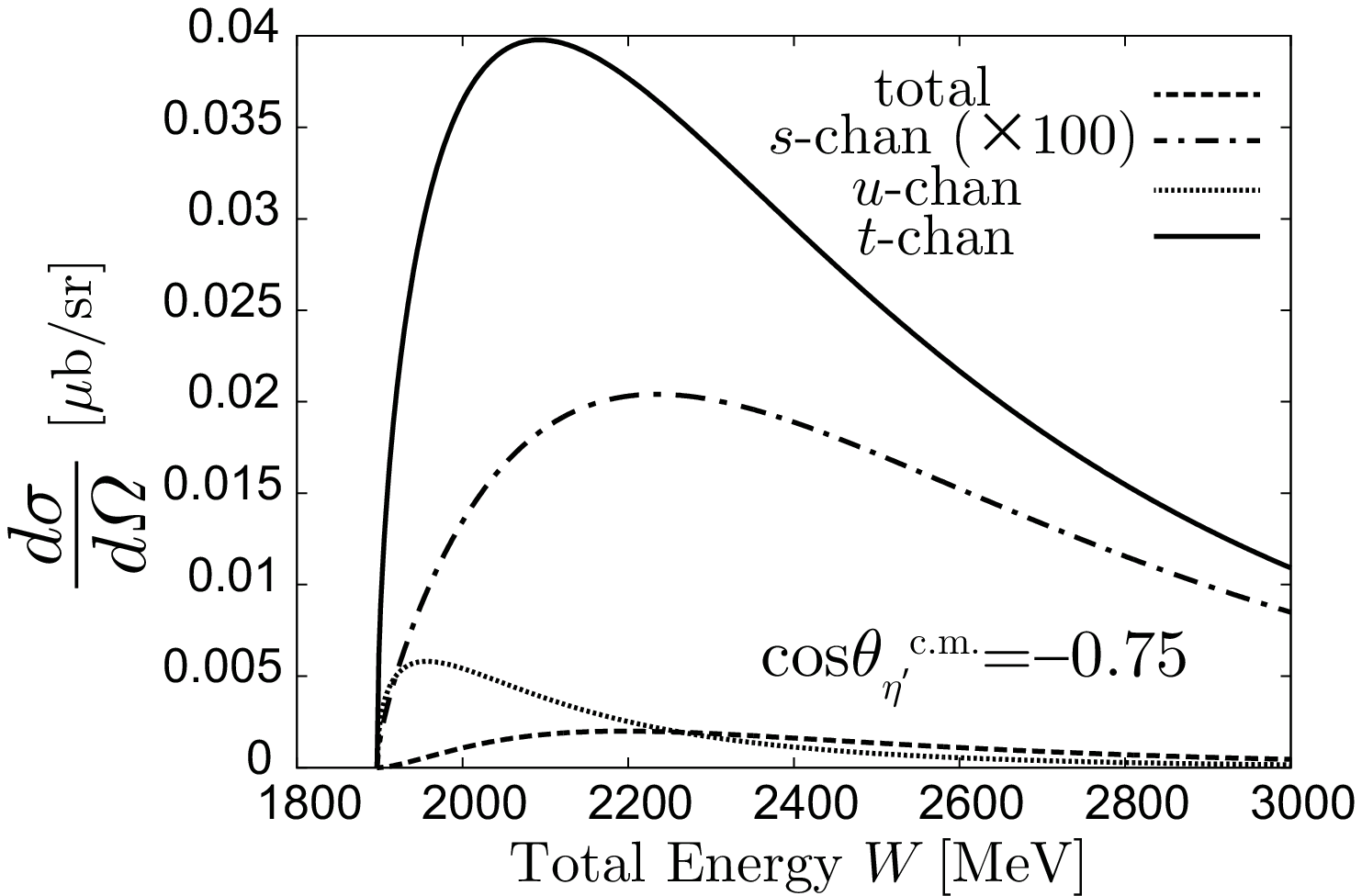}
 \end{minipage}
 \caption{Differential cross
 sections of the $\eta'$ photoproduction without FSI at $\cos\tetacm=0.75$ (left) and
 $-0.75$ (right) as functions of the total energy $W$ in the
 c.m. frame.
 In the left figure, $s$- and $u$-channel contributions are multiplied
 by factors $150$ and $5$, respectively, and the $s$-channel one in the
 right figure is multiplied by a factor $100$.
 \label{fig_diff_cs_detail}}
\end{figure}

Let us first discuss differential cross sections of the $\eta'$
photoproduction without the $\eta'N$ FSI as functions of the total
energy $W$ in the center-of-mass (c.m.) frame.
The results are shown in Fig.~\ref{fig_diff_cs_detail}.
The left and right panels of the figure correspond to the cases of an
$\eta'$ meson production at the forward angle $(\cos\tetacm=0.75)$ and
the backward one
$(\cos\tetacm=-0.75)$, respectively,
where $\tetacm$ denotes the angle between the
initial photon and the produced $\eta'$ meson in the c.m.
frame.
For the cutoff parameters,
we use $\Lambda=\Lambda_x=700$ MeV $(x=s,t,u)$.
In this figure, separate contributions from the $s$, $t$, and $u$
channels are plotted.
From the left panel of Fig.~\ref{fig_diff_cs_detail}, we find that
the cross section at the
forward angle is dominated by the $t$-channel contribution with the
vector-meson exchange.
On the other hand, the $u$-channel contribution of the second term of
Fig.~\ref{fig_diag_tree_photopro} has a large fraction at the backward
angle.

\begin{figure}[t]
 \centering
 \includegraphics[width=8cm]{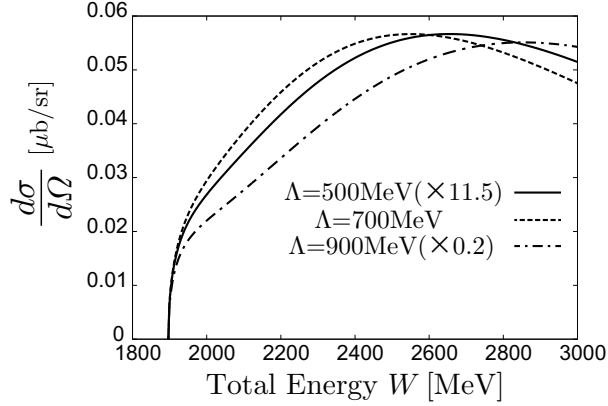}
 \caption{Cutoff dependence of
 the differential cross sections of the $\eta'$ photoproduction off
 the nucleon without the $\eta'N$ FSI.
 The cutoff parameter $\Lambda$ is varied as $\Lambda=500,$ $700,$ and
 $900$ MeV.
 Note that the results for $\Lambda=500$ and $900$ MeV are
 scaled by factors $11.5$ and $0.2$, respectively.
 \label{fig_cutoff_dep}}
\end{figure}

As is often the case, the reaction cross sections depend on the cutoff
parameters of the form factor.
Thus in Fig.~\ref{fig_cutoff_dep} we show the differential cross sections
at the  forward angle with varying the cutoff parameter
as $\Lambda=500$, $700$, and $900$ MeV without the $\eta'N$ FSI.
With the introduction of the form factor, some characteristic peak
structure may appear in the energy dependence of cross section due to
the competition of the increasing behavior of the phase space volume and
the decreasing behavior of the form factor as the energy (or the
relative momentum $q$) is increased.
The cross section is proportional to $q|F(q)|^2$, where $q$ is the
relative momentum of the final state $\eta'N$, and is related to the
kinetic energy $E$ by $E=q^2/2\mu$ in the non-relativistic approximation
for small $q$ ($\mu$ is the reduced mass).
By using the typical cutoff $\Lambda\sim 1$ GeV and $\mu\sim 0.5$ GeV
for the $\eta'N$ system, we find the peak position at around some
hundreds MeV above the threshold.
In Fig.~\ref{fig_cutoff_dep}, one finds that there is a peak
at $2.5-2.8$ GeV, that is, $600-900$ MeV above the $\eta'N$ threshold as
we expected, and that any characteristic structure around the threshold
does not appear with these cutoff parameters.

\begin{figure}[t]
 \begin{minipage}[t]{0.49\hsize}
  \centering
  \includegraphics[width=8cm]{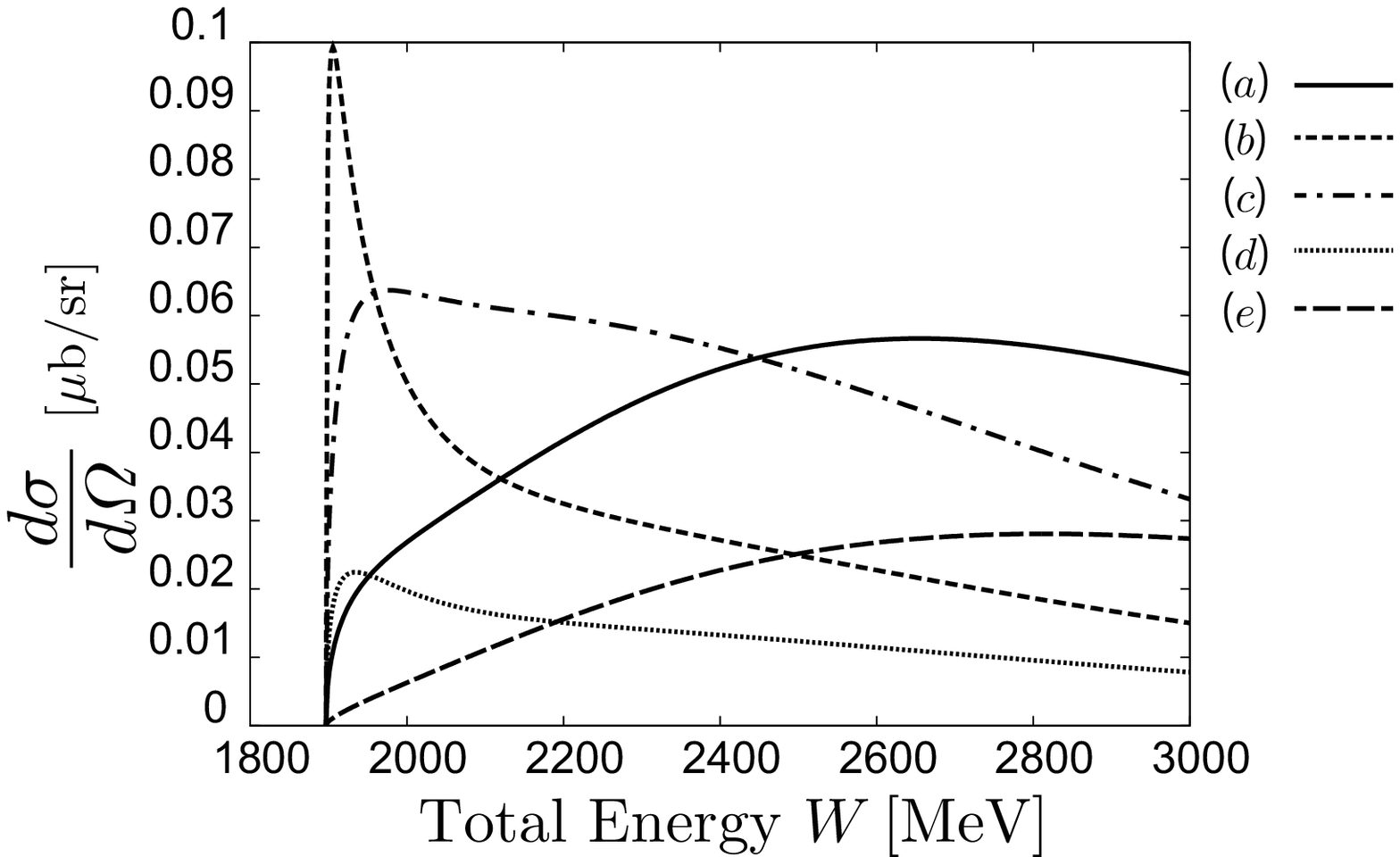}
 \end{minipage}
 \begin{minipage}[t]{0.49\hsize}
  \centering 
  \includegraphics[width=8cm]{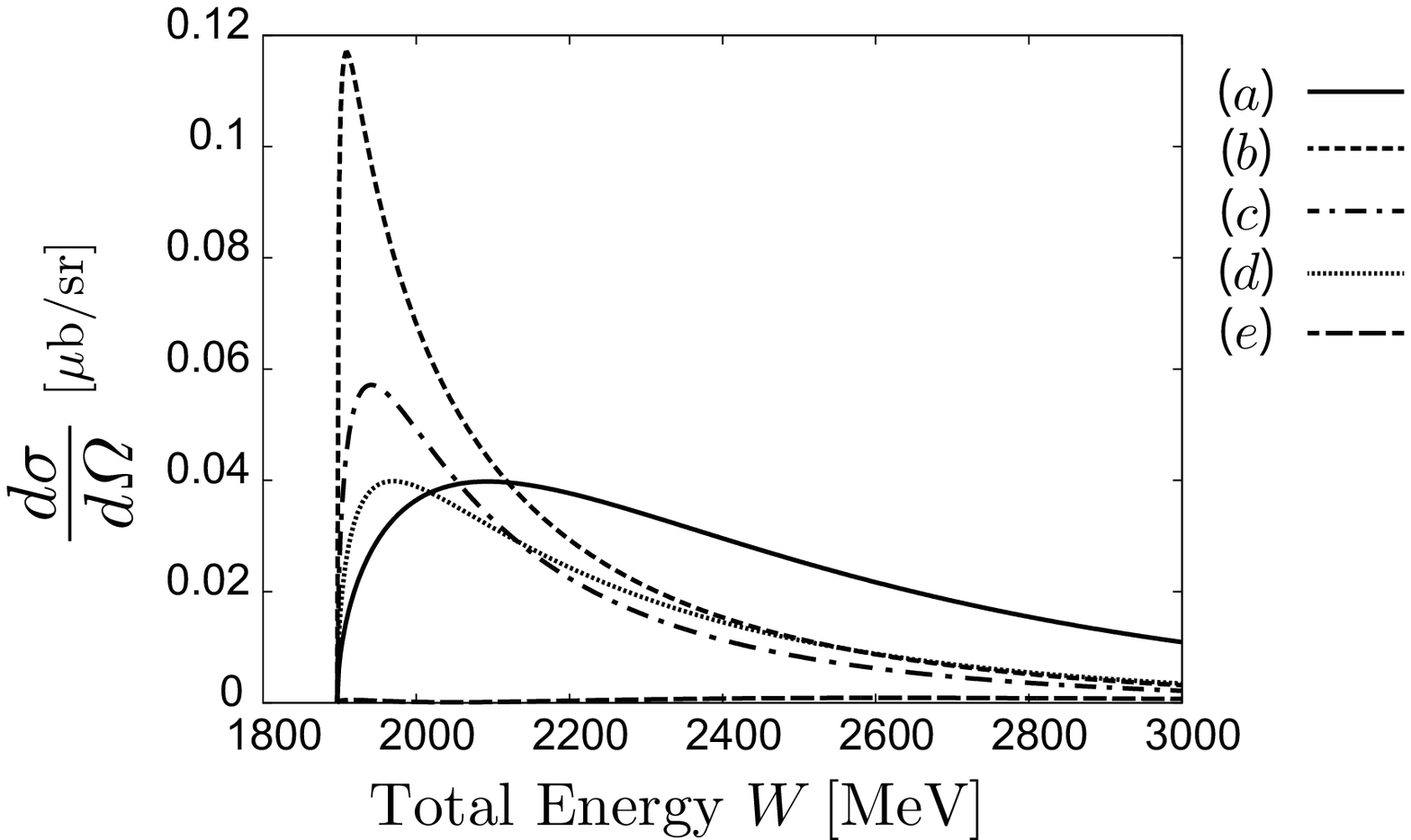}
 \end{minipage}
 \caption{Differential cross sections at the forward
 (left, $\cos\tetacm=0.75$) and the backward (right, $\cos\tetacm=-0.75$)
 angles as functions of $W$ with and without the $\eta'N$ FSI.
 The cases (a) to (e) in the legend follow those given in
 Table~\ref{tab_strength}.
 \label{fig_diff_cs_fsi_dep}}
\end{figure}

Now, Fig.~\ref{fig_diff_cs_fsi_dep} shows the total energy $W$
dependence of the differential cross sections with the inclusion of the
$\eta'N$ FSI.
The strength of the $\eta'N$ interaction is varied by changing the
parameter $g$ appearing in Eq.~(\ref{eq_eprimeN_kernel}) to see
the dependence on the strength of the $\eta'N$ FSI.
In the figure, the cases (a) to (e)
correspond to those explained in Sec.~\ref{subsec_epN_int};
(a) without FSI; (b), (c), (d) with attractive FSI; and (e)
with repulsive FSI.
Here, we use the same value for the cutoff parameter $\Lambda=700$ MeV in
all cases (a) to (e).
In this study, only the $\eta'N$ FSI in the $S$-wave part is included.
Therefore, we mainly focus on
the energy around the $\eta'N$ threshold in the following discussions.

In the left panel of Fig.~\ref{fig_diff_cs_fsi_dep} for the forward
production of the $\eta'$ meson,
we find a broad bump structure
around 2.6 GeV
for the case (a) without FSI,
which originates from the form factor as mentioned above.
With the inclusion of the $\eta'N$ FSI, the structure is modified:
In the case (b), a significant enhancement near the $\eta'N$ threshold
appears, which stems from the existence of a bound state just below the
threshold.
The enhancement becomes more moderate in the cases
(c) and (d), where there exists no bound state
around the $\eta'N$ threshold.
Thus, we find an enhancement of the forward cross
section near the $\eta'N$ threshold
due to
the attractive $\eta'N$ FSI.
In the case (e), where the $\eta'N$ FSI is repulsive,
one cannot find such an enhancement.
When the $\eta'$ meson is emitted at the backward angle, the
$\eta'N$ FSI gives similar effect on the energy dependence of the cross
sections as shown in the right panel of Fig.~\ref{fig_diff_cs_fsi_dep};
the larger cross sections near the $\eta'N$ threshold are obtained in the
cases (b), (c), and (d) than that in the case (a), and
one can see the suppression in the case (e) compared with the case (a).

\begin{figure}[t]
 \begin{minipage}[t]{0.49\hsize}
  \centering 
  \includegraphics[width=8cm]{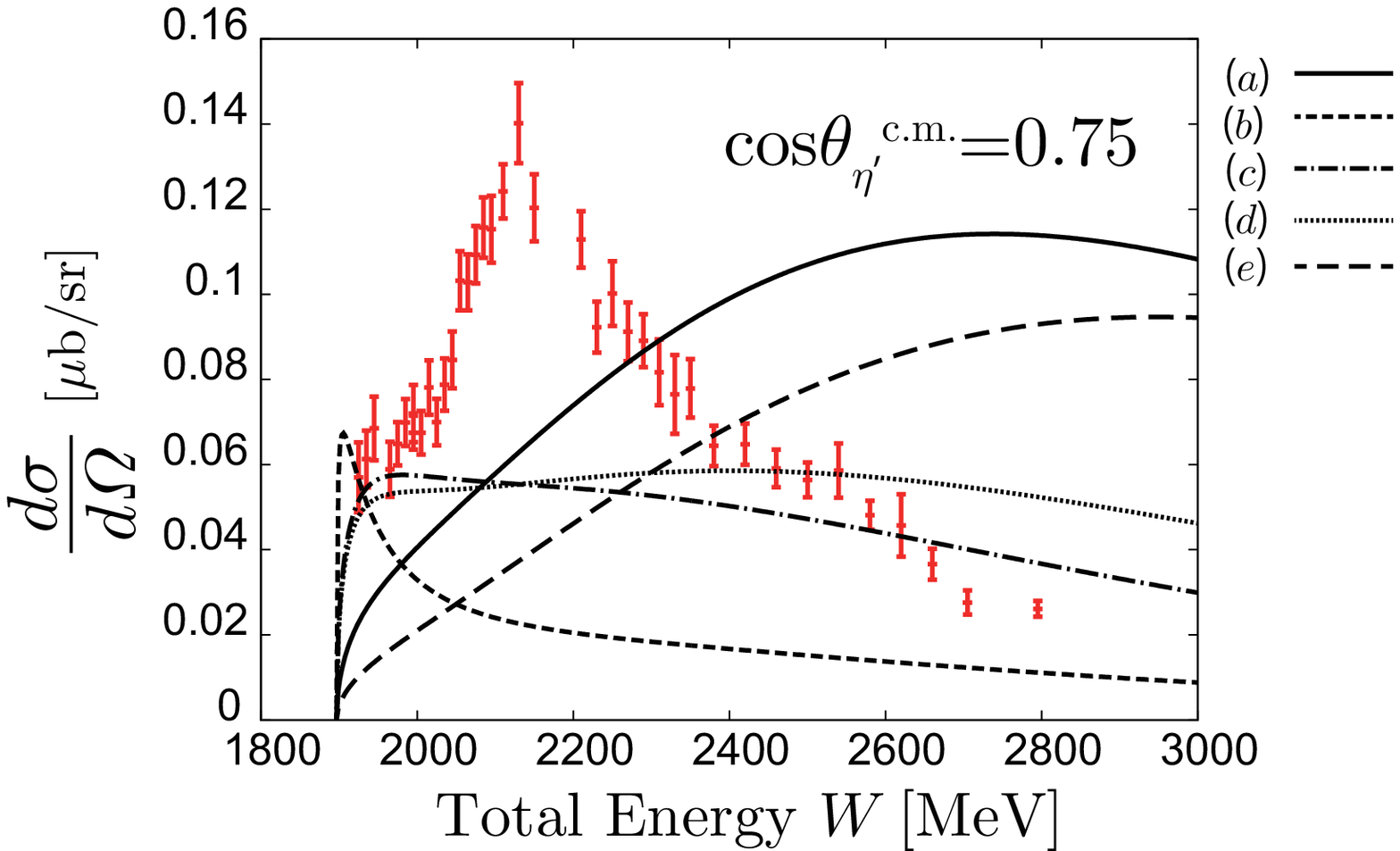}
 \end{minipage}
\begin{minipage}[t]{0.49\hsize}
 \centering 
 \includegraphics[width=8cm]{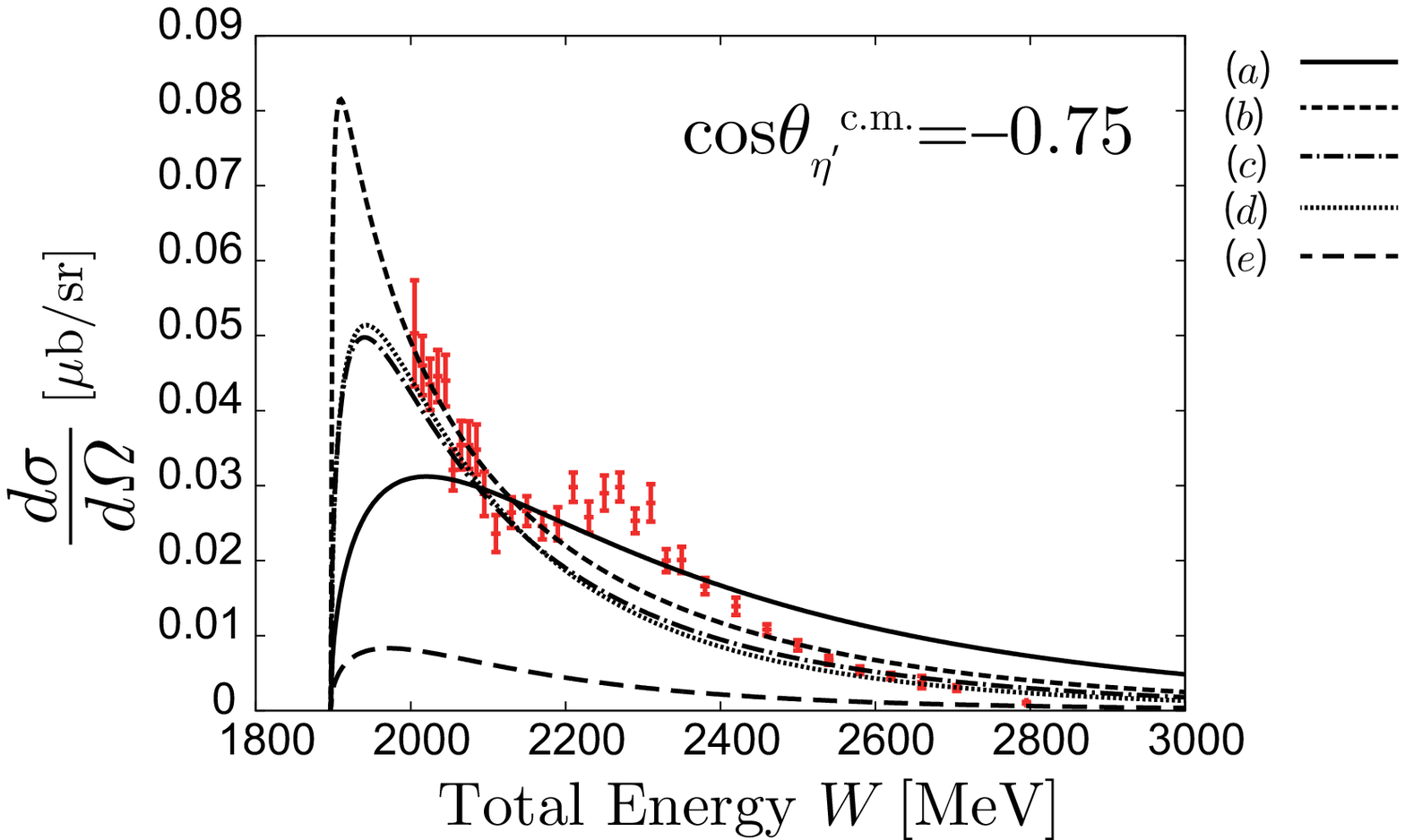}
\end{minipage}
 \caption{Differential cross sections of the $\eta'$ photoproduction at $\cos\tetacm=0.75$
 (left) and $-0.75$ (right) as functions of the total energy $W$.
 The cases (a) to (e) in the legend are the same as those in
 Fig.~\ref{fig_diff_cs_fsi_dep}.
 The points with the error bar are the experimental data taken from
 Ref.~\cite{Williams:2009yj}.}
 \label{fig_diff_cs_w_dep}
\end{figure}

In Fig.~\ref{fig_diff_cs_w_dep}, we show the result
of the differential cross sections
compared with the experimental data \cite{Williams:2009yj}.
In doing so, we have tuned the cutoff parameters $\Lambda_{u,s}$ and
$\Lambda_t$ for each strength of the $\eta'N$ FSI to make an optimal
comparison with the experimental data near the threshold at both forward
and backward angles.
The resulting cutoff parameters $\Lambda_x$ are summarized in
Table~\ref{tab_cutoff}.
\begin{table}[t]
 \centering
 \caption{Cutoff parameters $\Lambda_x$ used for the results shown in
 Fig.~\ref{fig_diff_cs_w_dep} in units of MeV.
 The cases (a) to (e) follow those of Table~\ref{tab_epn_scat}.}
 \label{tab_cutoff}
 \begin{ruledtabular}
  \begin{tabular}[t]{c|ccccc}
   &(a)&(b)&(c)&(d)&(e)\\\hline
   $\Lambda_{s,u}$&$600$ &$680$ &$680$ &$650$&$0$ \\ 
   $\Lambda_{t}$& $750$& $610$&$650$ &$790$&$840$ 
  \end{tabular}
 \end{ruledtabular}
\end{table}
At the forward angle shown in the left panel 
of Fig.~\ref{fig_diff_cs_w_dep}, the rapid increase
near the threshold is well reproduced in the cases (b), (c),
and (d), where the $\eta'N$ FSI is attractive, though
such behavior is not seen in the cases without the $\eta'N$ FSI, (a),
nor with the repulsive one, (e).
In the present method, we cannot reproduce a broad peak at around
$W=2.1$ GeV in the experimental data, which is considered to be due to a
resonance as discussed in Ref.~\cite{Nakayama:2005ts}.
In the present study, however, we do not consider
the resonance in that energy region, and we rather
focus our discussions on the near threshold behavior by the
$\eta'N$ FSI.
As we have mentioned before, there is no parameter set which
reproduces the
scattering length suggested by the COSY-11
experiment~\cite{Czerwinski:2014yot}.
We expect that the small scattering length leads to the
similar result to the case (a), where the effect of
the $\eta'N$ FSI is not taken into account and
the rapid increase near the $\eta'N$ threshold is not reproduced well.

Next, we move to the backward production of the $\eta'$ meson given in
the right panel of Fig.~\ref{fig_diff_cs_w_dep}.
Here, we note that the experimental data in Ref.~\cite{Williams:2009yj}
very near the $\eta'N$ threshold are missing, and
that only the data above 2 GeV are available.
We find a clear effect of the FSI at
the total energy $W$ below $2$ GeV.
The attractive $\eta'N$ FSI, the cases (b), (c), and (d),
leads to a rapid increase of the cross sections around the
$\eta'N$-threshold energy.
In the case (e),
the cross section is smaller than that in the case (a).
This difference of the cross sections near the $\eta'N$ threshold can
be a probe to investigate the strength of the low-energy $\eta'N$
interaction.
As corresponding to the broad peak seen in the experimental data at the
forward angle, a dip-like structure is seen at the backward angle at
the same energy region $W\sim 2.1$ GeV.
Once again, we do not discuss this structure because it may come from
the resonance effect as mentioned above.

\begin{figure}[t]
 \begin{minipage}[t]{0.49\hsize}
  \centering 
  \includegraphics[width=8cm]{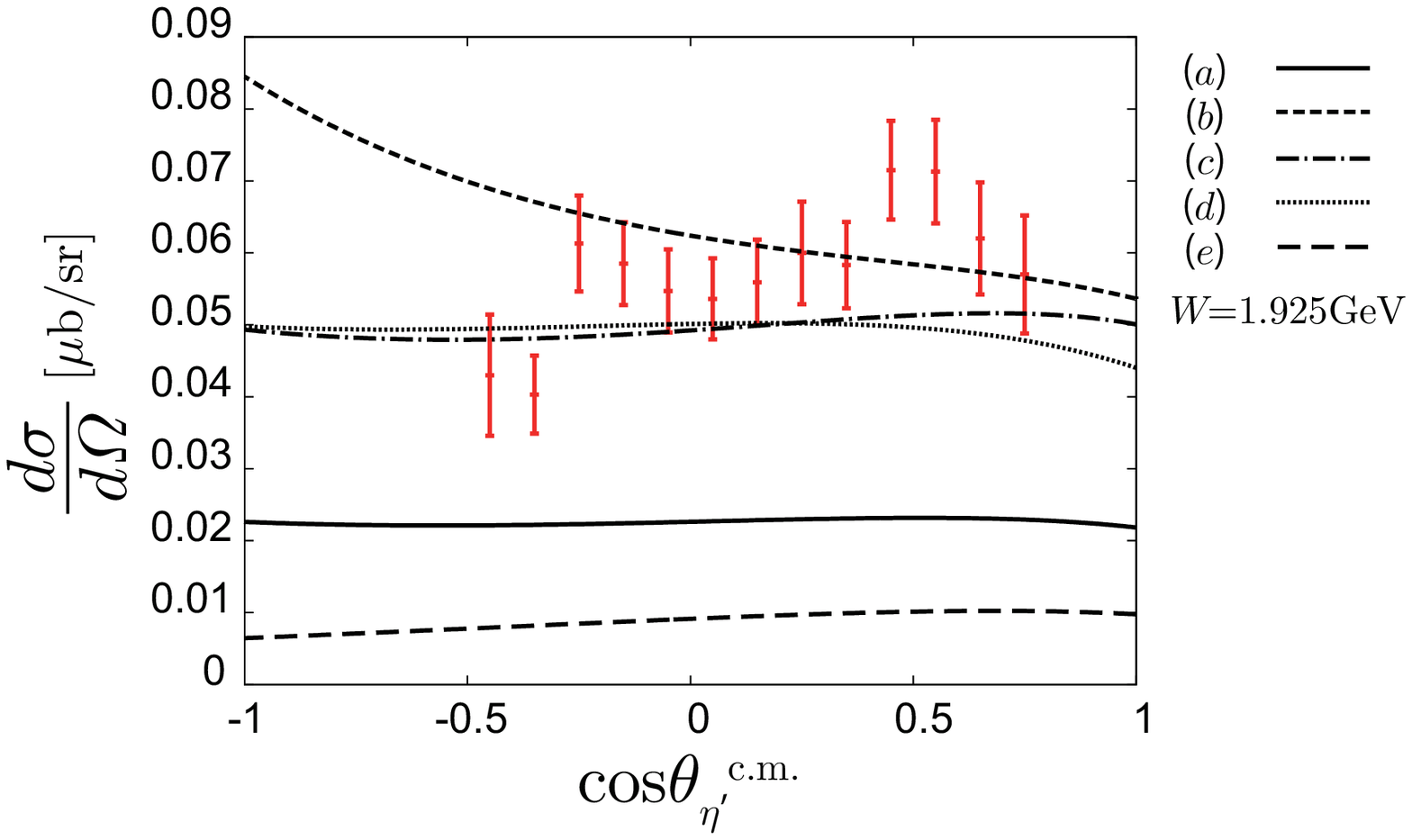}
 \end{minipage}
 \begin{minipage}[t]{0.49\hsize}
  \centering 
  \includegraphics[width=8cm]{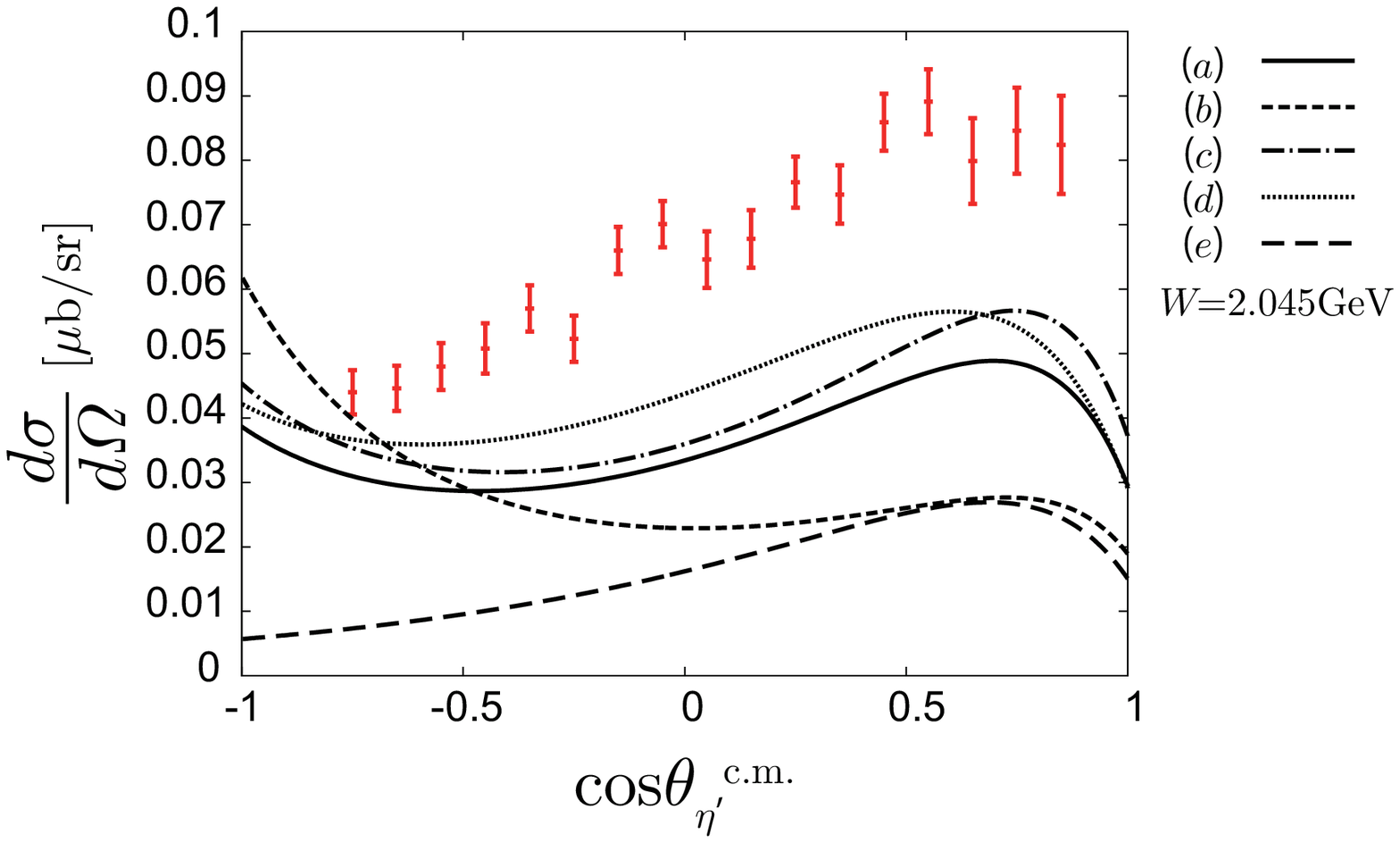}
 \end{minipage}
 \begin{minipage}[t]{0.49\hsize}
  \centering 
  \includegraphics[width=8cm]{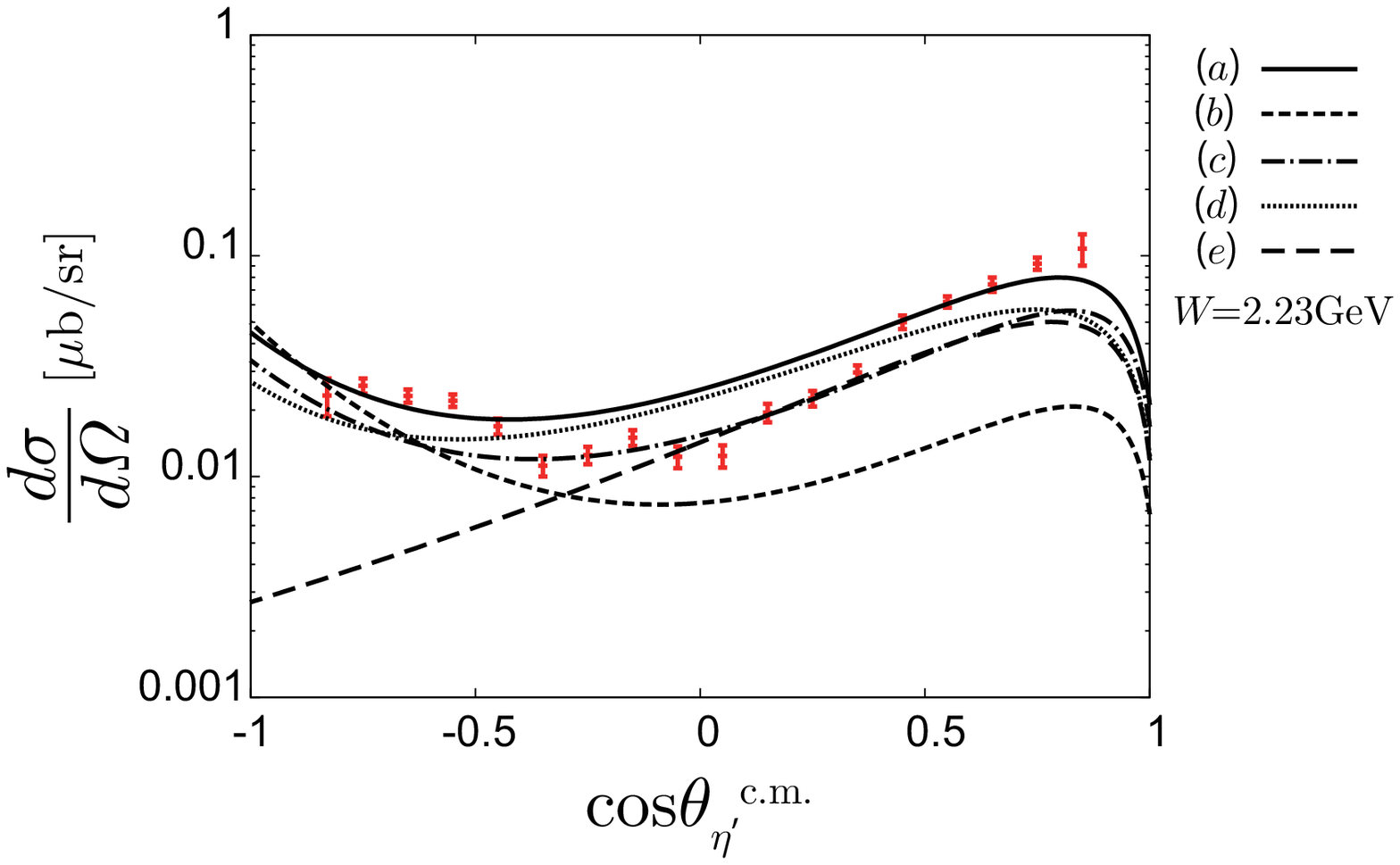}
 \end{minipage}
 \begin{minipage}[t]{0.49\hsize}
  \centering 
  \includegraphics[width=8cm]{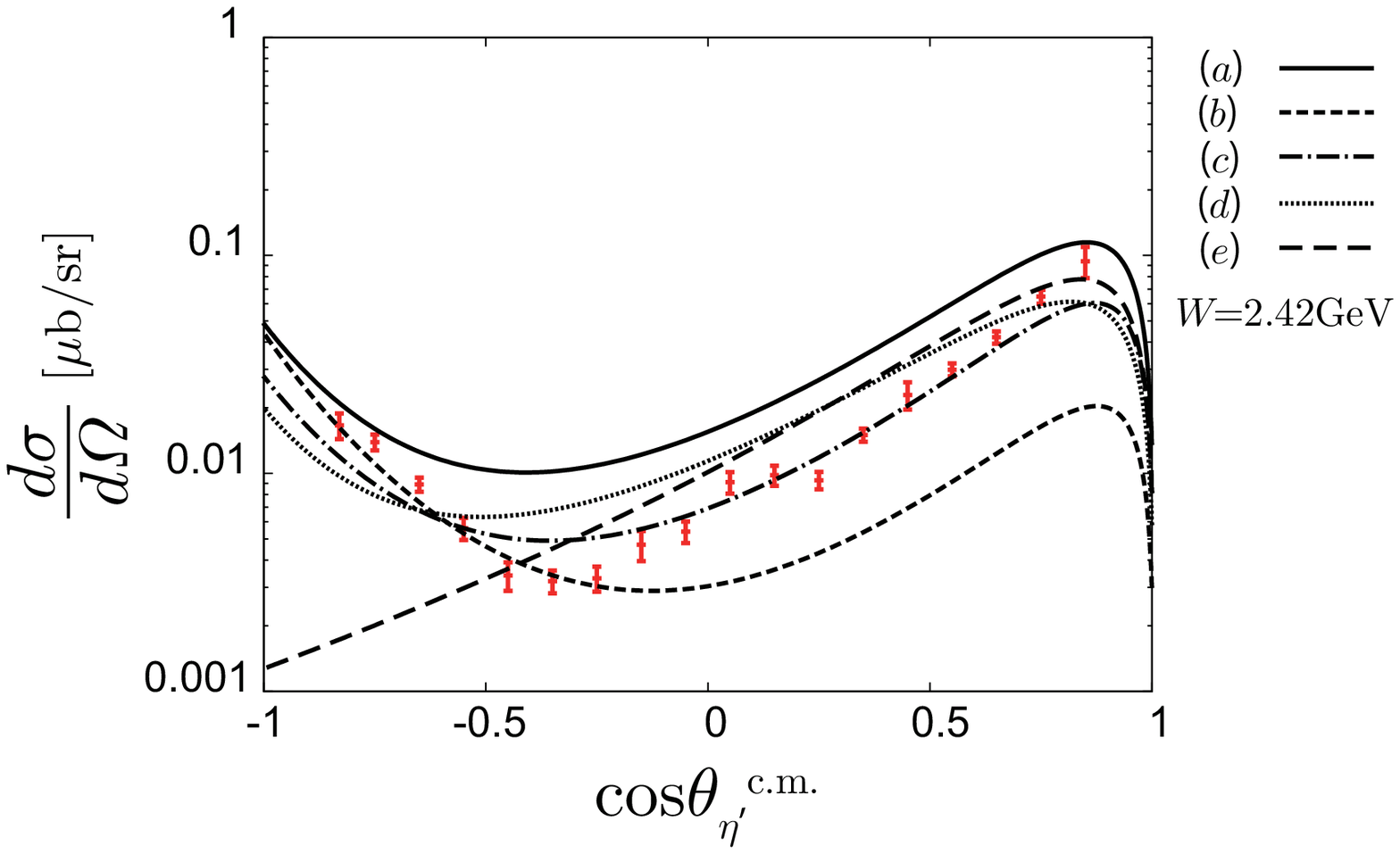}
 \end{minipage}
 \caption{Differential cross section of the $\eta'$ photoproduction as
 functions of $\cos\tetacm$ with and without the $\eta'N$ FSI at
 $W=1.925$ (upper left), $2.045$ (upper right), $2.230$ (lower left),
 and $2.420$ (lower right) GeV.
 The legend is the same as that in Fig.~\ref{fig_diff_cs_fsi_dep}.
 \label{fig_diff_cs_ct_dep}}
\end{figure}

The differential cross sections
at $W=1.925$, $2.045$, $2.230$, and $2.420$
GeV as functions of $\cos\tetacm$ are shown in
Fig.~\ref{fig_diff_cs_ct_dep}.
Around the $\eta'N$-threshold energy, $W=1.925$ GeV,
the cross sections of our calculation do not depend on the
variable $\cos\tetacm$ so much due to the expected $S$-wave dominance,
though the experimental data have some structure.
As we mentioned above, the
differences among the theoretical curves
come from those
in the strength of the $\eta'N$ FSI;
in the cases (b), (c), and (d)
which contain the attractive $\eta'N$ FSI,
the cross sections near the $\eta'N$ threshold have larger values
compared with that in the case (a).
At $W=2.045$ GeV and around $\cos\tetacm=1$,
there is discrepancy between our calculation and the experimental data.
This energy corresponds to the peak around $2.1$ GeV in the
experimental data in Fig.~\ref{fig_diff_cs_w_dep},
which may come from the resonance contribution as mentioned above.
At higher energies $(W=2.23$ and $2.42$ GeV$)$, the forward peak
structure stemming from the $t$-channel contribution
becomes more apparent.
The difference of the behavior at the backward angle
is caused by that of the $u$-channel contribution associated with the
change of the parameter $g$.
  
\begin{figure}[t]
 \centering 
 \includegraphics[width=8cm]{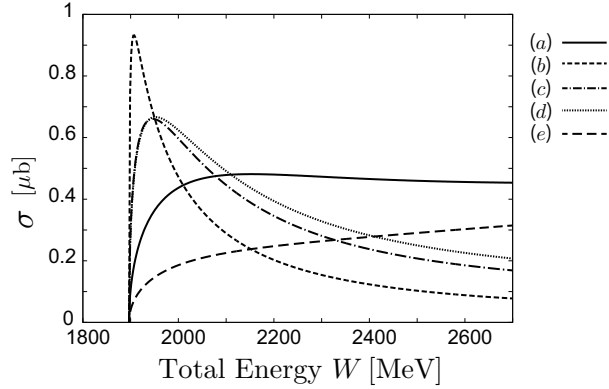}
 \caption{Total cross sections of
 the $\eta'$ photoproduction off the nucleon as functions of the total
 energy $W$.
 The legend is the same as that in Fig.~\ref{fig_diff_cs_fsi_dep}.
 \label{fig_total_cs}}
\end{figure}

In Fig.~\ref{fig_total_cs},
we show the total cross sections of the $\eta'$ photoproduction as
functions of the total energy $W$.
As in the case of the differential cross section, the enhancement of the
cross sections near the $\eta'N$ threshold is seen in the cases (b),
(c), and (d) with the attractive $\eta'N$ FSI.
In the case (e), the cross section is smaller than that in the case
(a).

Finally, we show the beam asymmetries $\Sigma$ against the scattering
angle $\tetacm$.
$\Sigma$ is defined as
\begin{align}
 \Sigma=\left.\left(\left.\frac{d\sigma}{d\Omega}\right|_{\phi=\pi/2}-\left.\frac{d\sigma}{d\Omega}\right|_{\phi=0}\right)\right/\left(\left.\frac{d\sigma}{d\Omega}\right|_{\phi=\pi/2}+\left.\frac{d\sigma}{d\Omega}\right|_{\phi=0}\right),  
\end{align}
where $\phi$ is the azimuthal angle from the polarization vector of the
photon in the initial state.
\begin{figure}
 \begin{minipage}[t]{0.49\hsize}
 \centering 
  \includegraphics[width=8cm]{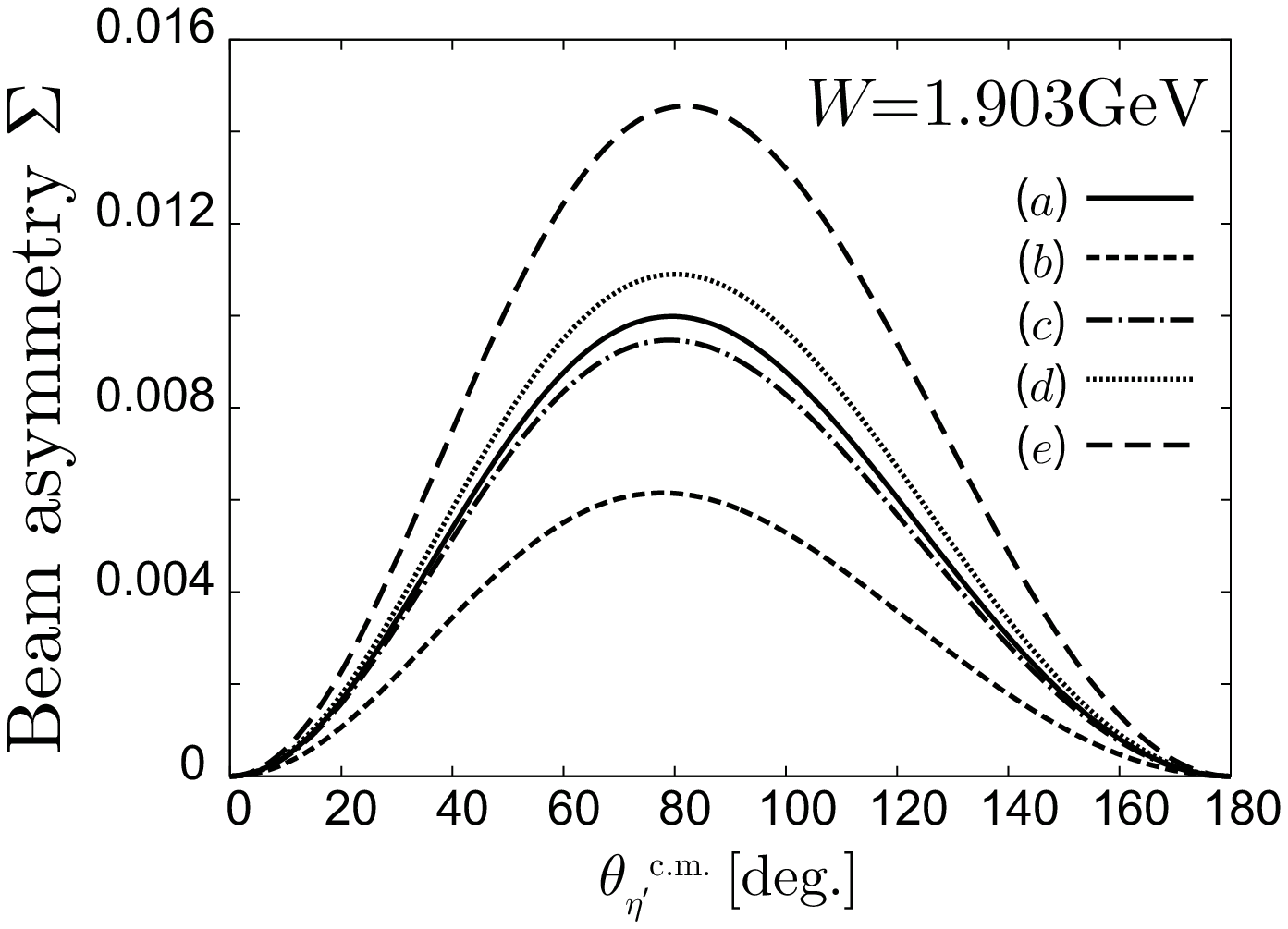}
 \end{minipage}
 \begin{minipage}[t]{0.49\hsize}
  \centering 
  \includegraphics[width=8cm]{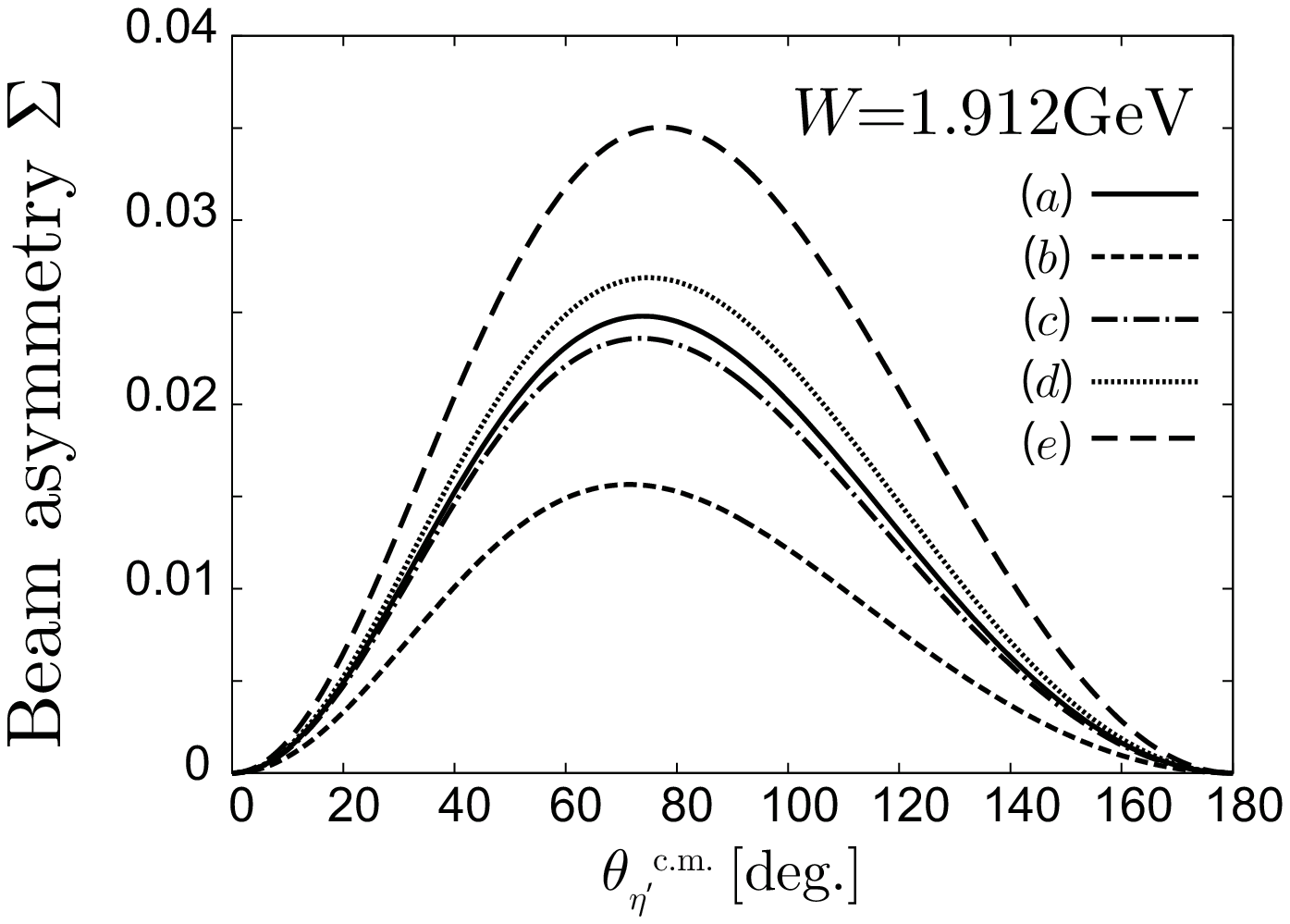} 
 \end{minipage}
 \caption{Beam asymmetries $\Sigma$ as functions of the scattering angle
 $\tetacm$ with the total energy $W=1.903$ (left) and $1.912$ (right) GeV.
 The cases (a) to (e) in the legend are the same as those in
 Fig.~\ref{fig_diff_cs_fsi_dep}.
 \label{fig_beam_asym_1}}
\end{figure}
The positive values of the beam asymmetries as shown in Fig.~\ref{fig_beam_asym_1} originate from the
dominant contribution of the $t$-channel diagram,
which is of the magnetic nature associated with the anomalous coupling
of $\gamma\eta'\rho$.
The behavior of the beam asymmetry is qualitatively different from
observed one \cite{Sandri:2014nqz}.
The difference may come from
the interference as pointed out in
Ref.~\cite{Sandri:2014nqz}.
Then, further development of the model, such as, the inclusion of
the higher partial-wave contribution may be necessary for the
description of the beam asymmetry.

\section{Summary and Outlook
\label{sec_summary}}
In this paper,
we investigated the $\eta'$ photoproduction off a nucleon with the
inclusion of the final-state interaction between the $\eta'$ meson and
nucleon based on the linear $\sigma$ model.
When there is an attractive final-state interaction, we found
an enhancement of the differential cross section
near the $\eta'N$ threshold, typically around or below $2$ GeV,
at the forward and backward angles $(\cos\tetacm=\pm0.75)$.
With an attractive $\eta'N$ interaction, the energy dependence of the
cross section near the $\eta'N$ threshold is reproduced fairly well.
Particularly, the magnitude of the enhancement near the threshold in
the backward production of the $\eta'$ meson seems
to be sensitive to the strength of the $\eta'N$ interaction.
The angular dependence of the differential cross section also
agrees with the experimental data in Ref.~\cite{Williams:2009yj}.
The enhancement around the $\eta'N$ threshold appears also in
the energy dependence of the total cross section.
Therefore, precise analysis of the threshold behavior is useful to
determine the $\eta'N$ interaction.
Despite these agreements,
the angular dependence of the beam
asymmetry shows qualitatively different behavior as in
the previous theoretical calculations~\cite{Sandri:2014nqz}.

The present study was based on a rather simple model and on the $S$-wave
scattering.
Other ingredients such as coupled channels of, e.g., $\eta N$ and $\pi
N$, higher partial waves, resonances, and so on, may be included.
These are expected to improve the aspects that cannot be explained in
the present study.

\begin{acknowledgments}
 This work is supported in part by the Grants-in-Aid for Science
 Research (C) by the JSPS (Grant Nos.~JP26400273 for A.~H. and
 JP26400275 for H.~N.).
\end{acknowledgments}


\end{document}